\documentclass[]{spie}  

\usepackage{amsmath,amsfonts,amssymb}
\usepackage{graphicx}
\usepackage[colorlinks=true, allcolors=blue]{hyperref}
\usepackage{units,subfig,libertine}
\usepackage{siunitx}

\title{Towards integrated superconducting detectors on lithium niobate waveguides}

\author[a]{Jan Philipp H{\"o}pker}
\author[a]{Moritz Bartnick}
\author[a]{Evan Meyer-Scott}
\author[a]{Frederik Thiele}
\author[a]{Stephan Krapick}
\author[a]{Nicola Montaut}
\author[a]{Matteo Santandrea}
\author[a]{Harald Herrmann}
\author[a]{Sebastian Lengeling}
\author[a]{Raimund Ricken}
\author[a]{Viktor Quiring}
\author[a]{Torsten Meier}
\author[b]{Adriana Lita}
\author[b]{Varun Verma}
\author[b]{Thomas Gerrits}
\author[b]{Sae Woo Nam}
\author[a]{Christine Silberhorn}
\author[a]{Tim J. Bartley}
\affil[a]{University of Paderborn, Warburger Str. 100, Paderborn, Germany}
\affil[b]{National Institute of Standards and Technology, 325 Broadway, Boulder, Colorado, USA}

\authorinfo{Send correspondence to Tim J. Bartley\\E-mail: tim.bartley@upb.de, Telephone: +49(0)5251 60 5881 }

\begin{document} 
\maketitle

\begin{abstract}
Superconducting detectors are now well-established tools for low-light optics, and in particular quantum optics, boasting high-efficiency, fast response and low noise. Similarly, lithium niobate is an important platform for  integrated optics given its high second-order nonlinearity, used for high-speed electro-optic modulation and polarization conversion, as well as frequency conversion and sources of quantum light. Combining these technologies addresses the requirements for a single platform capable of generating, manipulating and measuring quantum light in many degrees of freedom, in a compact and potentially scalable manner.
We will report on progress integrating tungsten transition-edge sensors (TESs) and amorphous tungsten silicide superconducting nanowire single-photon detectors (SNSPDs) on titanium in-diffused lithium niobate waveguides.
The travelling-wave design couples the evanescent field from the waveguides into the superconducting absorber. We will report on simulations and measurements of the absorption, which we can characterize at room temperature prior to cooling down the devices. Independently, we show how the detectors respond to flood illumination, normally incident on the devices, demonstrating their functionality.
\end{abstract}

\section{INTRODUCTION}
\label{sec:intro}  

Optical quantum technologies offer great potential to enhance computation, simulation, communication and measurement\cite{walmsley_quantum_2015}. Key to unlocking this potential is to develop hardware that preserves quantum coherence.  This comprises generating, manipulating and measuring quantum optical states with a high degree of efficiency and accuracy. Over recent years, developments in integrated quantum optics~\cite{tanzilli_genesis_2012} and superconducting single-photon detection~\cite{natarajan_superconducting_2012,hadfield_superconducting_2016} have made great strides in improving these tasks. Sources based on second-order nonlinear interactions in waveguides have proven to be very efficient, enabling amongst many other examples,  high purity~\cite{harder_optimized_2013}, bright quantum states~\cite{harder_single-mode_2016}, and cascaded quantum processes~\cite{krapick_-chip_2016}. Similarly, manipulating quantum states in waveguides has enabled landmark progress in, for example, tests of computationally hard sampling problems~\cite{broome_photonic_2013,spring_boson_2013,tillmann_experimental_2013,crespi_integrated_2013}, reconfigurable computational circuitry~\cite{carolan_universal_2015}, quantum memories~\cite{saglamyurek_broadband_2011} and nonlinear frequency conversion~\cite{de_greve_quantum-dot_2012}. 

Integrating detectors in a travelling wave design~\cite{hu_efficiently_2009} on waveguide structures has also garnered high levels of interest. There exist a broad range of superconducting detector materials and optical integration platforms, as summarized in Table~\ref{tab:WGDet}. The advantage of lithium niobate in this context is the connection between well-established technology in the classical optical domain~\cite{sohler_integrated_2008}, such as high-speed switching, phase modulations, and high-efficiency frequency conversion, with high-efficiency single-photon detectors.  
The challenge with lithium niobate has been successful material bonding and saturation of the internal detection efficiency of the superconductor.  While functional detectors have been deposited on lithium niobate as a substrate~\cite{tanner_superconducting_2012}, saturation of the internal detection efficiency and deposition directly onto waveguide structures has not yet been demonstrated.

\begin{table}[ht]
\caption{Summary of detector types integrated onto waveguides} 
\label{tab:WGDet}
\begin{center}       
\begin{tabular}{|l|c|c|c|c|c|} 
\hline
\rule[-1ex]{0pt}{3.5ex}   & NbN & NbTiN & WSi & MoSi  & W (TES)\\
\hline
\rule[-1ex]{0pt}{3.5ex} GaAs  &\cite{sprengers_waveguide_2011,jahanmirinejad_photon-number_2012,reithmaier_-chip_2013,sahin_waveguide_2013,zhou_superconducting_2014,kaniber_integrated_2016,najafi_-chip_2015,mattioli_photon-counting_2016} & & &\cite{li_nano-optical_2016}  &\\
\hline
\rule[-1ex]{0pt}{3.5ex} Si  &\cite{pernice_high-speed_2012} &\cite{akhlaghi_waveguide_2015} & &  &\\
\hline
\rule[-1ex]{0pt}{3.5ex} Si$_3$Ni$_4$ &\cite{cavalier_light_2011,ferrari_waveguide-integrated_2015,kahl_waveguide_2015} &\cite{schuck_nbtin_2013,schuck_waveguide_2013,schuck_quantum_2016} & \cite{beyer_waveguide-coupled_2015,shainline_room-temperature-deposited_2017}&  &\\
\hline
\rule[-1ex]{0pt}{3.5ex} Diamond &\cite{rath_superconducting_2015}&\cite{atikian_superconducting_2014}& &  &\\
\hline
\rule[-1ex]{0pt}{3.5ex} AlN &\cite{najafi_-chip_2015}&& &&  \\
\hline
\rule[-1ex]{0pt}{3.5ex} Au (plasmonic) &\cite{heeres_-chip_2010,heeres_quantum_2013}&& &&  \\
\hline
\rule[-1ex]{0pt}{3.5ex} LiNbO$_3$ &\cite{tanner_superconducting_2012}&& This work& & This work \\
\hline
\rule[-1ex]{0pt}{3.5ex} SiO$_2$&&& &  &\cite{gerrits_-chip_2011,calkins_high_2013}\\
\hline
\end{tabular}
\end{center}
\end{table}

Here, we present progress towards integrating functional superconducting detectors on titanium in-diffused waveguide in lithium niobate.  
The absorbers we consider are superconducting nanowire single-photon detectors (SNSPDs) and transition-edge sensors (TESs). The SNSPDs are constructed from thin-film amorphous tungsten silicide (a-WSi) patterned into a meander structure, which can produce very high efficiency when fiber-coupled~\cite{marsili_detecting_2013}. The amorphous nature means bonding to the lithium niobate is relatively straightforward; there is no need to lattice-match between the two materials. The TES devices are made from superconducting tungsten thin film and have also been demonstrated with high-efficiency~\cite{lita_counting_2008} when fiber coupled.

In this paper, we present initial results demonstrating the successful bonding of the detector structures onto the waveguide regions. We show simulations of the expected absorption of these devices, as well as experimental results from additional waveguide loss caused by these detectors (which places an upper bound on their efficiency). Finally, we present results showing the optical response of the devices under flood illumination. 

\section{Waveguide fabrication and detector deposition}
\begin{figure}[hbt]
\centering
\includegraphics[width=0.6\linewidth]{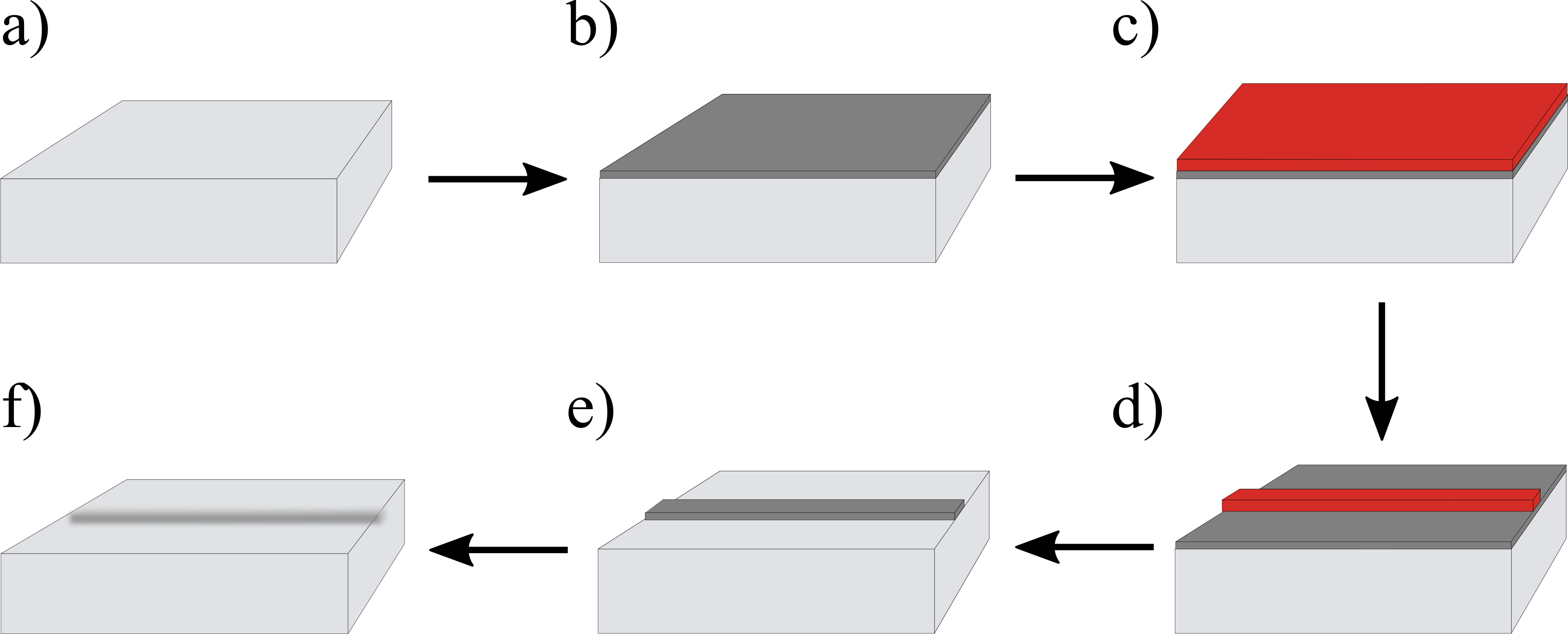}
\caption{\label{fig01} Waveguide fabrication scheme: a) Bare sample, b) Titanium deposition, c) Deposited positive photoresist, d) Patterned photoresist stripe, e) Remaining Ti stripe after etching and resist removal, f) In-diffused Ti waveguide.}
\end{figure}

Our integrated wave-guiding structures are based on the diffusion of titanium into congruent lithium niobate (LN). As depicted in the processing scheme Fig. \ref{fig01}, we first deposit a thin film of titanium of $\approx$\SI{80}{\nano\meter} thickness onto the LN substrate using electron-beam evaporation. In order to pattern the titanium film, positive photoresist is spin-coated on top and soft-baked afterwards. Using vacuum contact lithography, 
we produces homogeneous resist stripes of \SI{5}{\micro\meter}, \SI{6}{\micro\meter}, and \SI{7}{\micro\meter} in width that act as a protective layer for the subsequent wet etching of titanium. After removing residual photoresist, the remaining Ti stripes are diffused into the LN substrate. Due to the locally increased refractive index in the titanium-diffused regions we have fabricated structures that are capable of guiding light at wavelengths around \SI{1550}{\nano\meter} in single spatial modes in both polarizations.

To create superconducting nanowire single photon detectors (SNSPDs) on top of our waveguide structures, we deposit planar thin films of amorphous tungsten silicide (\SI{3.5}{\nano\meter}) and silicon (\SI{2.5}{\nano\meter}) using DC magnetron sputtering. 
The nanowire meanders are then patterned in a \SI{20}{\micro\meter}$\times$\SI{20}{\micro\meter}square area using electron-beam lithography.  These are coupled to contact pads and wirebonded to electrical connectors. 

The transition-edge sensor (TES) devices are DC-sputtered tungsten thin films of \SI{20}{\nano\meter} thickness. We investigated both \SI{10}{\micro\meter}$\times$\SI{10}{\micro\meter} and \SI{25}{\micro\meter}$\times$\SI{25}{\micro\meter} devices. These devices were deposited directly on top of the waveguides, whereby the \SI{10}{\micro\meter}$\times$\SI{10}{\micro\meter} device also included \SI{100}{\micro\meter}$\times$\SI{4}{\micro\meter} fins on either side of the detector to increase the overall coupling of the evanescent field to the detector. 
Note that in a traveling-wave geometry we can freely deposit multiple detectors on a single waveguide, which can significantly increase the overall absorption of the device.

\section{Mode profiles and expected absorption}
\begin{figure}[ht]
    \centering
		\subfloat[\label{fig:layout}
	]{\includegraphics[width=0.2\linewidth]{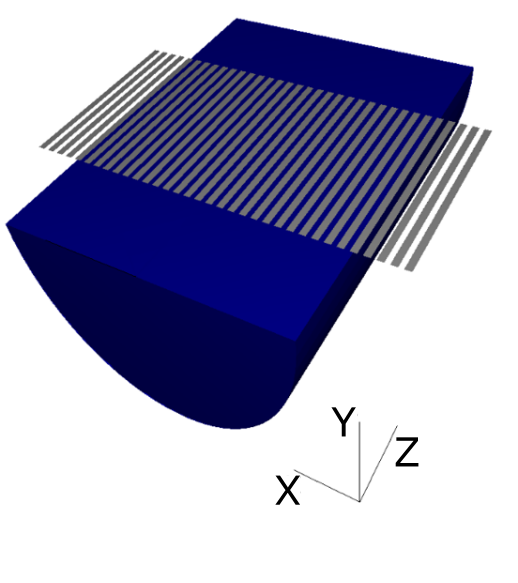}}\hfill
		\subfloat[ \label{fig:IndProf}
		]{\includegraphics[width=0.38\linewidth]{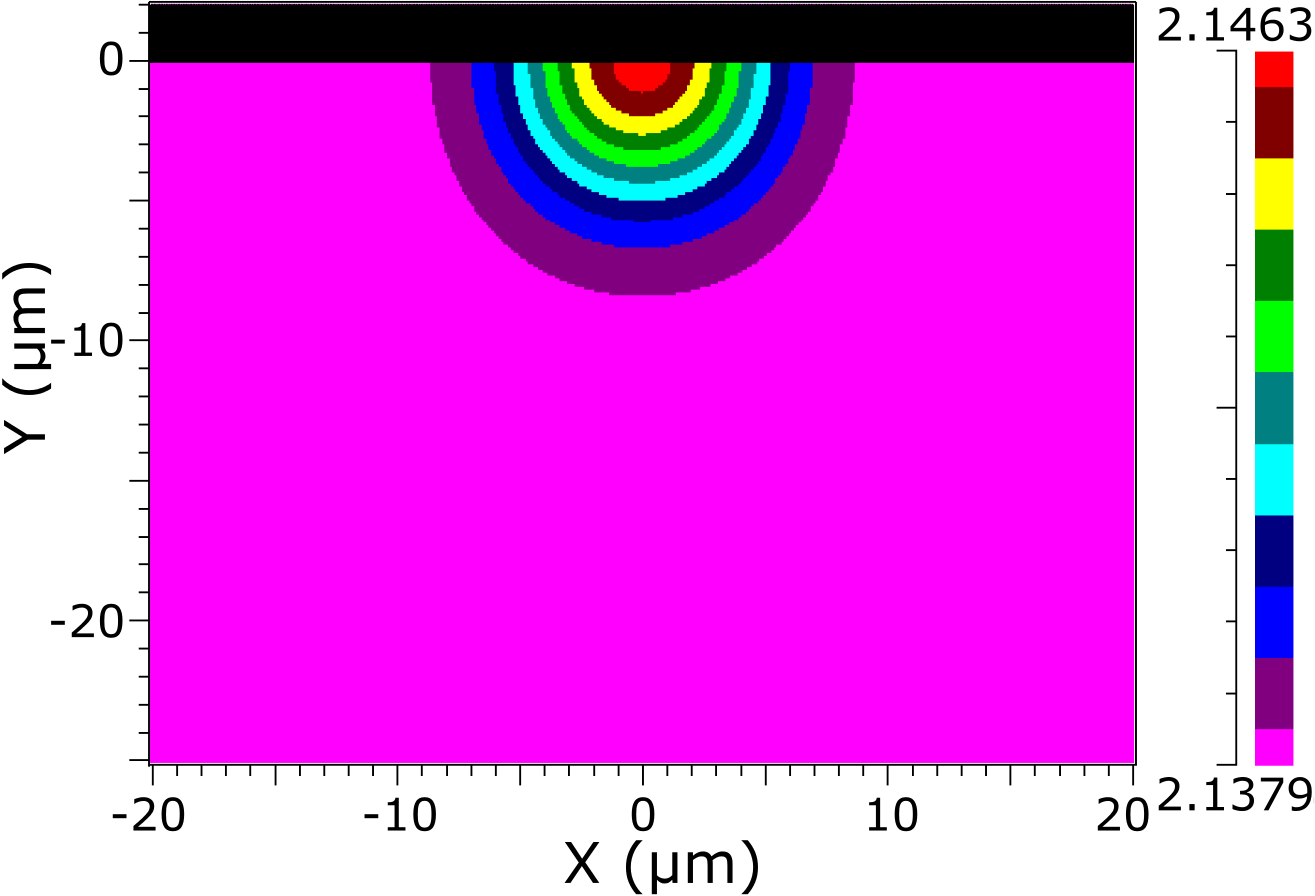}}\hfill
		\subfloat[\label{fig:ModeProf}
		]{\includegraphics[width=0.38\linewidth]{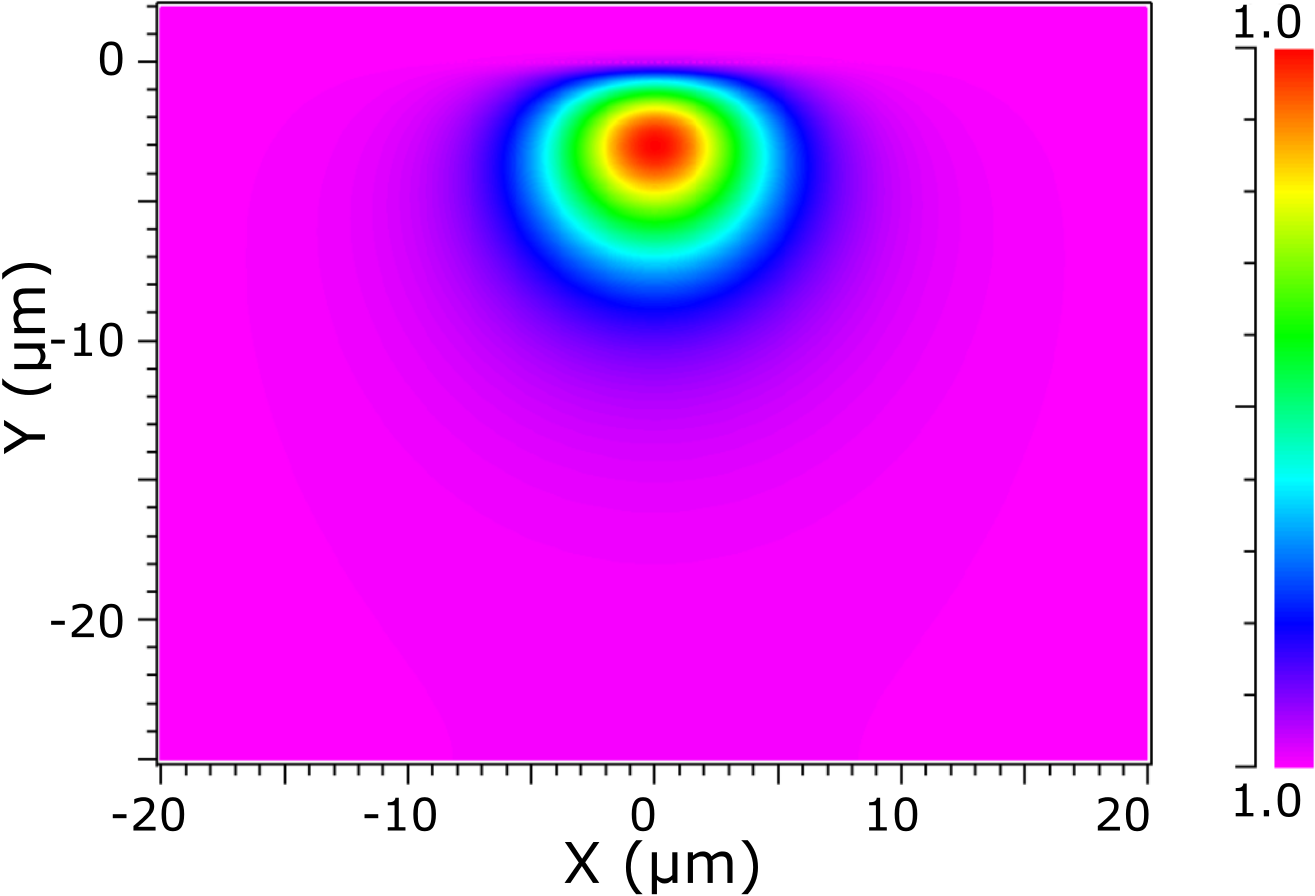}}\hfill
				
				\caption{Modeling absorption and mode structures. (a) Modeled SNSPD structure on top of waveguide. (b) Index profile of Ti in-diffused LN waveguides. (c) Resulting mode profile. The absorption can be inferred from the imaginary part of the effective index.}\label{fig:Sims}
\end{figure}

To investigate the expected coupling from the evanescent field of the waveguide mode, we use commercial mode solvers to calculate the effect of the detector.  Using the imaginary part of the refractive index of the absorbing material, we can calculate the expected absorption of this mode, and therefore the absorption over the length of the detector material.  As an example, we consider the SNSPD structure shown in Fig.~\ref{fig:layout}. We begin with the index profile of the waveguides shown in Fig.~\ref{fig:IndProf}. The profile is based on published models for the ordinary~\cite{edwards_temperature-dependent_1984} and extraordinary~\cite{jundt_temperature-dependent_1997} refractive indices. The resulting TM mode is shown in  Fig.~\ref{fig:ModeProf}. From the imaginary part of the effective index, we expect that our \SI{20}{\micro\meter}$\times$\SI{20}{\micro\meter} detector structure should give rise to about  $\approx$0.25~\% overall absorption, depending on the precise layer thickness following deposition. By contrast, the TES structure, comprising a \SI{10}{\micro\meter}$\times$\SI{10}{\micro\meter}$\times$\SI{0.02}{\micro\meter} volume of tungsten with two \SI{100}{\micro\meter}$\times$\SI{4}{\micro\meter}$\times$\SI{0.02}{\micro\meter} fins projecting from the central detector core, is expected to absorb between 11~\% and 18~\% of the incoming light, depending on the width of the waveguides. These mode profiles assume a negligible change to the mode structure as the beam propagates along the waveguide. 

\section{Loss measurements}
Experimentally, we can infer an upper limit on the absorption of the detectors by investigating the additional waveguide loss they induce. To characterize the linear loss of the waveguides, we use the interferometric technique of Regener and Sohler~\cite{regener_loss_1985}. This method exploits the Fresnel reflection at the interfaces of the waveguide to build a low-finesse Fabry-Perot interferometer. The mode reflectivity is given by
\begin{equation}
R=\left(\frac{1-n_\textrm{eff}}{1+n_\textrm{eff}}\right)^2
\end{equation}
where $n_\textrm{eff}$ is the effective refractive index of the mode. We take $n_\textrm{eff,TE}=2.212$ as the index of the transverse electric (TE) mode, and $n_\textrm{eff,TM}=2.139$ for the transverse magnetic (TM) mode. 
By varying the temperature of the sample and therefore the effective length of the interferometer, the interference pattern is scanned. From the visibility of the resulting interference pattern $V=\tfrac{I_\textrm{max}-I_\textrm{min}}{I_\textrm{max}+I_\textrm{min}}$ and the sample length $L$, one can calculate the loss $\alpha$ in units of dB/cm according to~\cite{regener_loss_1985}
\begin{equation}
\alpha=\frac{4.34}{L}\left(\ln R-\ln \frac{V}{2}\right)~.
\end{equation}
We calculate the losses of waveguides of width \SI{5}{\micro\meter}, \SI{6}{\micro\meter}, \SI{7}{\micro\meter}, for samples with and without detector structures on the same chip, before and after the detectors were deposited. This allows us to discriminate additional losses due to the deposition process itself, from the desired detector absorption by comparing waveguides with detectors and waveguides without detectors which have undergone the same processing steps. We carry out this characterisation for each polarisation, determined by a polarizer placed prior to incoupling. This procedure relies on precisely polished end faces perpendicular to the waveguide, such that only the Fresnel reflection need be considered.

\begin{figure}[thb]
\centering
\includegraphics[width=0.6\linewidth]{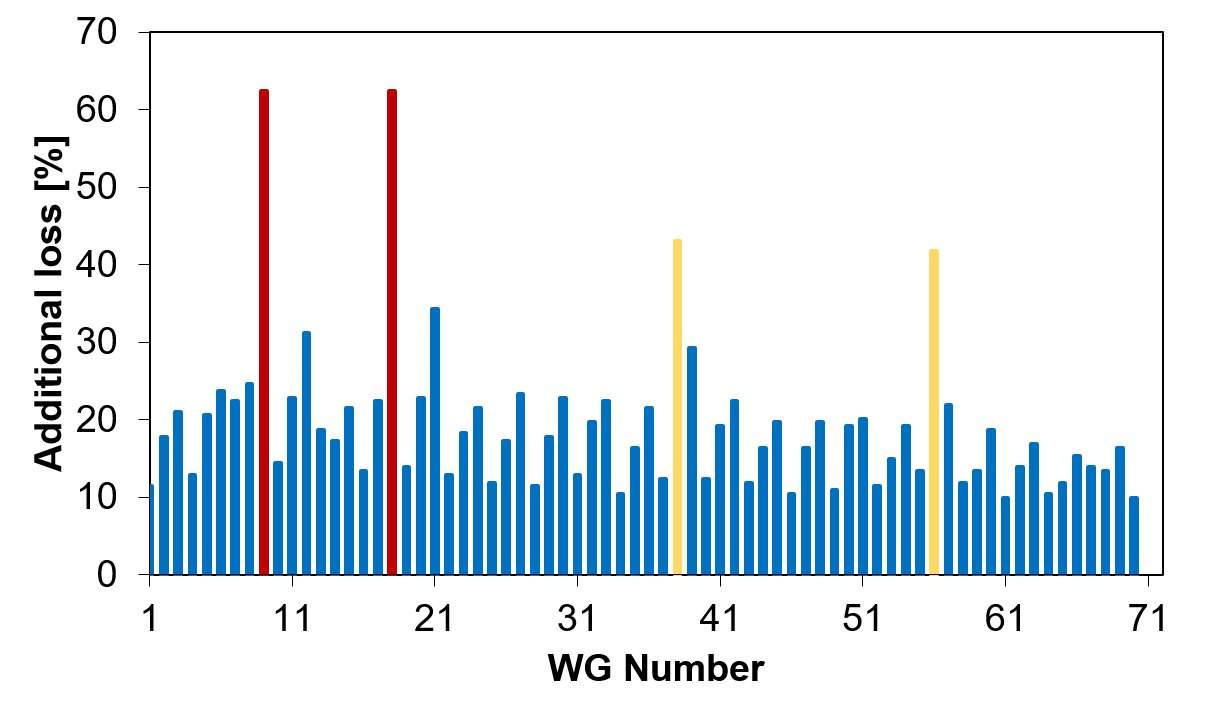}
\caption{\label{fig:TESloss} Additional losses for each waveguide after TES deposition: five detectors on waveguides 9 and 18 (marked in red) and three detectors on waveguides 38 and 56 (marked in gold). The rest of the waveguides have no deposited detectors. The periodicity in additional loss reflects the periodicity in waveguide widths of \SI{5}{\micro\meter}, \SI{6}{\micro\meter} and \SI{7}{\micro\meter}.}
\end{figure}

An example of the loss measurements for the TM mode is shown in Fig.~\ref{fig:TESloss}. In this case, 5 TES detectors are deposited on waveguides 9 and 18, and 3 detectors were placed on waveguides 38 and 56. All other waveguides had no detector structures placed on top; any additional loss arises from the chip processing. The losses caused by these detector structures are clear, and we deduce that each detector on a induces $11.3\pm1.0~\%$ and $13.5^{+1.4}_{-1.8}~\%$ absorption in the TM  mode, on waveguides of width \SI{6}{\micro\meter} and \SI{7}{\micro\meter}, respectively. Note that the uncertainty is dominated by systematic errors in determining the losses, rather than the statistics of the measurement. These values agree reasonably well with the simulated absorption values of 15.1~\% and 17.8~\% in each case respectively. For the TE mode, the additional loss corresponding to detector absorption was measured to be $0.9\pm0.7\%$, which agrees with the simulated absorption of 1~\%.

\section{Flood illumination tests}
\begin{figure}[htb]
    \centering
   \subfloat[\label{fig:SNSPD_sat}
	]{
        \includegraphics[width=0.45\linewidth]{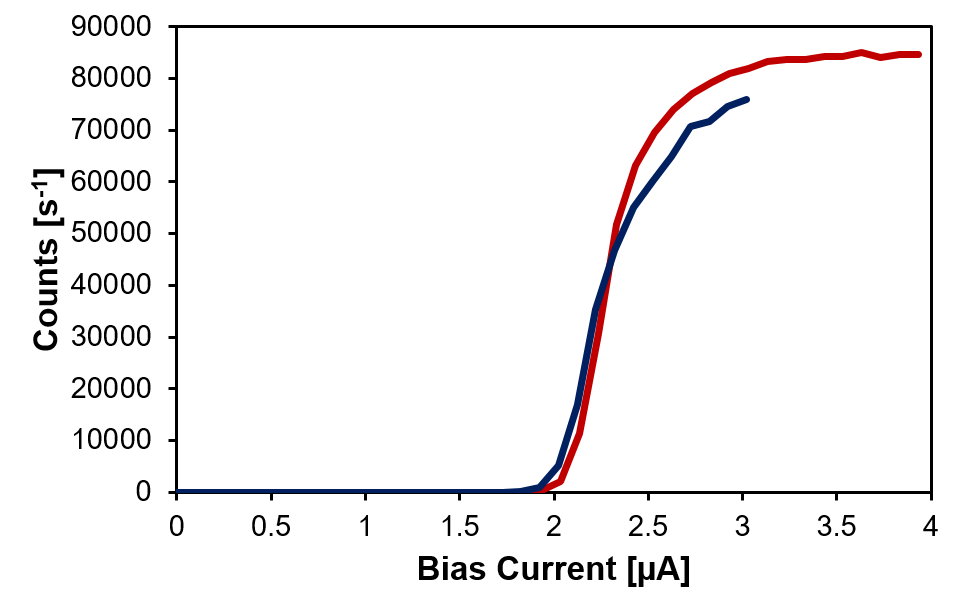}}\hspace{0.7cm}
   \subfloat[ \label{fig:TES_flood}
		]{
        \includegraphics[width=0.35\linewidth]{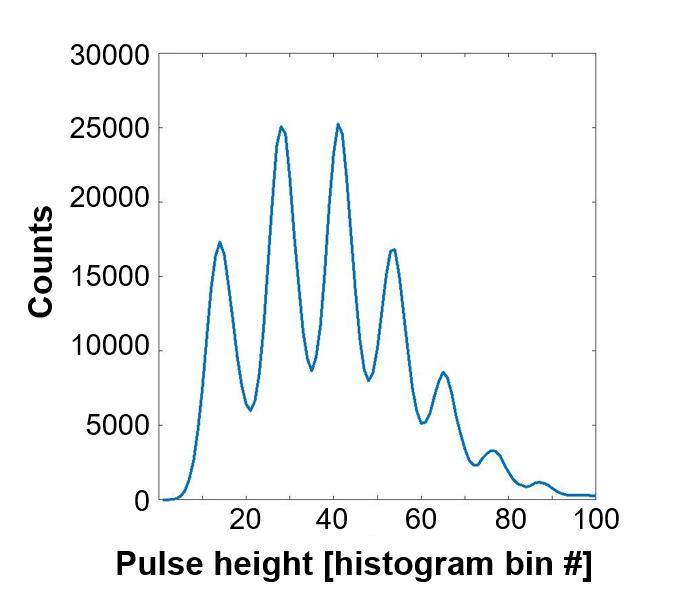}}\hfill
				\caption{Response of superconducting detectors deposited on Ti in-diffused LN waveguides under flood illumination. (a) Count rate as a function of bias current for two SPSPD devices. (b) Histogram of TES pulse heights in response to flood illumination. The different peaks correspond to different numbers of photons.}\label{fig:flood}
\end{figure}

In order to verify that the detector structures are compatible with the LN substrate, their optical response was tested under flood illumination. Of the six SNSPD devices tested, two superconducted correctly and responded to the incoming light, as shown in Fig.~\ref{fig:SNSPD_sat}. Of the two functioning devices, one clearly shows a plateau at bias currents above \SI{3}{\micro\ampere}, a clear signature of saturation of the internal detection efficiency. The second approaches a plateau before it reaches its critical current. A critical parameter in improving the yield for these devices is to minimize the surface roughness of the substrate. 

In Fig.~\ref{fig:TES_flood}, we show a histogram of the peak heights of the TES response. The separated peaks correspond to the energy absorbed by different numbers of photons. The data was collected from a \SI{25}{\micro\meter}$\times$\SI{25}{\micro\meter} TES without fins. The critical temperature ($T_c$) of this device was \SI{165}{\milli\kelvin}. The photon number resolution is slightly reduced as compared to TESs fabricated on silicon\cite{lita_counting_2008}. Further investigation should reveal whether the thermal link between the detector and the underlying lithium niobate substrate behaves differently than with the standard silicon substrate. However, the response is clear and the characteristic photon-number resolution is also easily identified.

\section{Conclusion}
We have shown a number of techniques used to characterize the absorption of thin-film substrates designed for use as integrated superconducting detectors on titanium in-diffused lithium niobate waveguides. Furthermore, we have verified the optical response of the devices under flood illumination, indicating that integrated detectors on lithium niobate are possible. Initial modeling and measurement show there is substantial optimization to be undertaken to maximize the coupling from the evanescent fields of these waveguides to the absorber. This may involve additional layers of dielectrics to modify the mode structure, or modifying the waveguide fabrication recipe to achieve waveguides with a larger evanescent field. Similarly, there is substantial progress required to identify the optimal operating conditions for the devices, for example optimizing their geometry with respect to the waveguides as well as their electrical and thermal behavior.

\acknowledgments 
 T. J. B. acknowledges financial support from the DFG (Deutsche Forschungsgemeinschaft) through the SFB/TRR 142.
This is a contribution of NIST, an agency of the U.S. government, not subject to copyright

\bibliography{SPIE_final} 

\begin{thebibliography}{10}

\bibitem{walmsley_quantum_2015}
Walmsley, I.~A., ``Quantum optics: {Science} and technology in a new light,''
  {\em Science}~{\bf 348},  525--530 (May 2015).

\bibitem{tanzilli_genesis_2012}
Tanzilli, S., Martin, A., Kaiser, F., De~Micheli, M., Alibart, O., and
  Ostrowsky, D., ``On the genesis and evolution of {Integrated} {Quantum}
  {Optics},'' {\em Laser \& Photon. Rev.}~{\bf 6},  115--143 (Jan. 2012).

\bibitem{natarajan_superconducting_2012}
Natarajan, C.~M., Tanner, M.~G., and Hadfield, R.~H., ``Superconducting
  nanowire single-photon detectors: physics and applications,'' {\em
  Superconductor Science and Technology}~{\bf 25},  063001 (June 2012).

\bibitem{hadfield_superconducting_2016}
Hadfield, R.~H. and Johansson, G.,  [{\em Superconducting {Devices} in
  {Quantum} {Optics}}{\nolinebreak\hspace{0.1em}]}, Springer (Feb. 2016).

\bibitem{harder_optimized_2013}
Harder, G., Ansari, V., Brecht, B., Dirmeier, T., Marquardt, C., and
  Silberhorn, C., ``An optimized photon pair source for quantum circuits,''
  {\em Opt. Express}~{\bf 21},  13975--13985 (June 2013).

\bibitem{harder_single-mode_2016}
Harder, G., Bartley, T.~J., Lita, A.~E., Nam, S.~W., Gerrits, T., and
  Silberhorn, C., ``Single-mode parametric-down-conversion states with 50
  photons as a source for mesoscopic quantum optics,'' {\em Phys. Rev.
  Lett.}~{\bf 116},  143601 (Apr 2016).

\bibitem{krapick_-chip_2016}
Krapick, S., Brecht, B., Herrmann, H., Quiring, V., and Silberhorn, C.,
  ``On-chip generation of photon-triplet states,'' {\em Opt. Express}~{\bf 24},
   2836--2849 (Feb. 2016).

\bibitem{broome_photonic_2013}
Broome, M.~A., Fedrizzi, A., Rahimi-Keshari, S., Dove, J., Aaronson, S., Ralph,
  T.~C., and White, A.~G., ``Photonic {Boson} {Sampling} in a {Tunable}
  {Circuit},'' {\em Science}~{\bf 339},  794--798 (Feb. 2013).

\bibitem{spring_boson_2013}
Spring, J.~B., Metcalf, B.~J., Humphreys, P.~C., Kolthammer, W.~S., Jin, X.-M.,
  Barbieri, M., Datta, A., Thomas-Peter, N., Langford, N.~K., Kundys, D.,
  Gates, J.~C., Smith, B.~J., Smith, P. G.~R., and Walmsley, I.~A., ``Boson
  {Sampling} on a {Photonic} {Chip},'' {\em Science}~{\bf 339},  798--801 (Feb.
  2013).

\bibitem{tillmann_experimental_2013}
Tillmann, M., Daki{\'c}, B., Heilmann, R., Nolte, S., Szameit, A., and Walther,
  P., ``Experimental boson sampling,'' {\em Nat Photon}~{\bf 7},  540--544
  (July 2013).

\bibitem{crespi_integrated_2013}
Crespi, A., Osellame, R., Ramponi, R., Brod, D.~J., Galv{\~a}o, E.~F.,
  Spagnolo, N., Vitelli, C., Maiorino, E., Mataloni, P., and Sciarrino, F.,
  ``Integrated multimode interferometers with arbitrary designs for photonic
  boson sampling,'' {\em Nat Photon}~{\bf 7},  545--549 (July 2013).

\bibitem{carolan_universal_2015}
Carolan, J., Harrold, C., Sparrow, C., Mart{\'i}n-L{\'o}pez, E., Russell,
  N.~J., Silverstone, J.~W., Shadbolt, P.~J., Matsuda, N., Oguma, M., Itoh, M.,
  Marshall, G.~D., Thompson, M.~G., Matthews, J. C.~F., Hashimoto, T., O'Brien,
  J.~L., and Laing, A., ``Universal linear optics,'' {\em Science}~{\bf 349},
  711--716 (Aug. 2015).

\bibitem{saglamyurek_broadband_2011}
Saglamyurek, E., Sinclair, N., Jin, J., Slater, J.~A., Oblak, D.,
  Bussi{\`e}res, F., George, M., Ricken, R., Sohler, W., and Tittel, W.,
  ``Broadband waveguide quantum memory for entangled photons,'' {\em
  Nature}~{\bf 469},  512--515 (Jan. 2011).

\bibitem{de_greve_quantum-dot_2012}
De~Greve, K., Yu, L., McMahon, P.~L., Pelc, J.~S., Natarajan, C.~M., Kim,
  N.~Y., Abe, E., Maier, S., Schneider, C., Kamp, M., H{\"o}fling, S.,
  Hadfield, R.~H., Forchel, A., Fejer, M.~M., and Yamamoto, Y., ``Quantum-dot
  spin-photon entanglement via frequency downconversion to telecom
  wavelength,'' {\em Nature}~{\bf 491},  421--425 (Nov. 2012).

\bibitem{hu_efficiently_2009}
Hu, X., Holzwarth, C.~W., Masciarelli, D., Dauler, E.~A., and Berggren, K.~K.,
  ``Efficiently {Coupling} {Light} to {Superconducting} {Nanowire}
  {Single}-{Photon} {Detectors},'' {\em IEEE Transactions on Applied
  Superconductivity}~{\bf 19},  336--340 (June 2009).

\bibitem{sohler_integrated_2008}
Sohler, W., Hu, H., Ricken, R., Quiring, V., Vannahme, C., Herrmann, H.,
  B{\"u}chter, D., Reza, S., Grundk{\"o}tter, W., Orlov, S., Suche, H.,
  Nouroozi, R., and Min, Y., ``Integrated {Optical} {Devices} in {Lithium}
  {Niobate},'' {\em Optics \& Photonics News, OPN}~{\bf 19},  24--31 (Jan.
  2008).

\bibitem{tanner_superconducting_2012}
Tanner, M.~G., Alvarez, L. S.~E., Jiang, W., Warburton, R.~J., Barber, Z.~H.,
  and Hadfield, R.~H., ``A superconducting nanowire single photon detector on
  lithium niobate,'' {\em Nanotechnology}~{\bf 23}(50),  505201 (2012).

\bibitem{sprengers_waveguide_2011}
Sprengers, J.~P., Gaggero, A., Sahin, D., Jahanmirinejad, S., Frucci, G.,
  Mattioli, F., Leoni, R., Beetz, J., Lermer, M., Kamp, M., H{\"o}fling, S.,
  Sanjines, R., and Fiore, A., ``Waveguide superconducting single-photon
  detectors for integrated quantum photonic circuits,'' {\em Appl. Phys.
  Lett.}~{\bf 99},  181110 (Oct. 2011).

\bibitem{jahanmirinejad_photon-number_2012}
Jahanmirinejad, S., Frucci, G., Mattioli, F., Sahin, D., Gaggero, A., Leoni,
  R., and Fiore, A., ``Photon-number resolving detector based on a series array
  of superconducting nanowires,'' {\em Appl. Phys. Lett.}~{\bf 101},  072602
  (Aug. 2012).

\bibitem{reithmaier_-chip_2013}
Reithmaier, G., Lichtmannecker, S., Reichert, T., Hasch, P., M{\"u}ller, K.,
  Bichler, M., Gross, R., and Finley, J.~J., ``On-chip time resolved detection
  of quantum dot emission using integrated superconducting single photon
  detectors,'' {\em Scientific Reports}~{\bf 3},  01901 (May 2013).

\bibitem{sahin_waveguide_2013}
Sahin, D., Gaggero, A., Zhou, Z., Jahanmirinejad, S., Mattioli, F., Leoni, R.,
  Beetz, J., Lermer, M., Kamp, M., H{\"o}fling, S., and Fiore, A., ``Waveguide
  photon-number-resolving detectors for quantum photonic integrated circuits,''
  {\em Appl. Phys. Lett.}~{\bf 103},  111116 (Sept. 2013).

\bibitem{zhou_superconducting_2014}
Zhou, Z., Jahanmirinejad, S., Mattioli, F., Sahin, D., Frucci, G., Gaggero, A.,
  Leoni, R., and Fiore, A., ``Superconducting series nanowire detector counting
  up to twelve photons,'' {\em Opt. Express}~{\bf 22},  3475--3489 (Feb. 2014).

\bibitem{kaniber_integrated_2016}
Kaniber, M., Flassig, F., Reithmaier, G., Gross, R., and Finley, J.~J.,
  ``Integrated superconducting detectors on semiconductors for quantum optics
  applications,'' {\em Appl. Phys. B}~{\bf 122},  115 (May 2016).

\bibitem{najafi_-chip_2015}
Najafi, F., Mower, J., Harris, N.~C., Bellei, F., Dane, A., Lee, C., Hu, X.,
  Kharel, P., Marsili, F., Assefa, S., Berggren, K.~K., and Englund, D.,
  ``On-chip detection of non-classical light by scalable integration of
  single-photon detectors,'' {\em Nature Communications}~{\bf 6},  6873 (Jan.
  2015).

\bibitem{mattioli_photon-counting_2016}
Mattioli, F., Zhou, Z., Gaggero, A., Gaudio, R., Leoni, R., and Fiore, A.,
  ``Photon-counting and analog operation of a 24-pixel photon number resolving
  detector based on superconducting nanowires,'' {\em Opt. Express}~{\bf 24},
  9067--9076 (Apr. 2016).

\bibitem{li_nano-optical_2016}
Li, J., Kirkwood, R.~A., Baker, L.~J., Bosworth, D., Erotokritou, K., Banerjee,
  A., Heath, R.~M., Natarajan, C.~M., Barber, Z.~H., Sorel, M., and Hadfield,
  R.~H., ``Nano-optical single-photon response mapping of waveguide integrated
  molybdenum silicide ({MoSi}) superconducting nanowires,'' {\em Opt.
  Express}~{\bf 24},  13931--13938 (June 2016).

\bibitem{pernice_high-speed_2012}
Pernice, W. H.~P., Schuck, C., Minaeva, O., Li, M., Goltsman, G.~N., Sergienko,
  A.~V., and Tang, H.~X., ``High-speed and high-efficiency travelling wave
  single-photon detectors embedded in nanophotonic circuits,'' {\em Nature
  Communications}~{\bf 3},  1325 (Dec. 2012).

\bibitem{akhlaghi_waveguide_2015}
Akhlaghi, M.~K., Schelew, E., and Young, J.~F., ``Waveguide integrated
  superconducting single-photon detectors implemented as near-perfect absorbers
  of coherent radiation,'' {\em Nature Communications}~{\bf 6},  9233 (Sept.
  2015).

\bibitem{cavalier_light_2011}
Cavalier, P., Vill{\'e}gier, J.-C., Feautrier, P., Constancias, C., and Morand,
  A., ``Light interference detection on-chip by integrated {SNSPD} counters,''
  {\em AIP Advances}~{\bf 1},  042120 (Oct. 2011).

\bibitem{ferrari_waveguide-integrated_2015}
Ferrari, S., Kahl, O., Kovalyuk, V., Goltsman, G.~N., Korneev, A., and Pernice,
  W. H.~P., ``Waveguide-integrated single- and multi-photon detection at
  telecom wavelengths using superconducting nanowires,'' {\em Appl. Phys.
  Lett.}~{\bf 106},  151101 (Apr. 2015).

\bibitem{kahl_waveguide_2015}
Kahl, O., Ferrari, S., Kovalyuk, V., Goltsman, G.~N., Korneev, A., and Pernice,
  W. H.~P., ``Waveguide integrated superconducting single-photon detectors with
  high internal quantum efficiency at telecom wavelengths,'' {\em Scientific
  Reports}~{\bf 5},  10941 (June 2015).

\bibitem{schuck_nbtin_2013}
Schuck, C., Pernice, W. H.~P., and Tang, H.~X., ``{NbTiN} superconducting
  nanowire detectors for visible and telecom wavelengths single photon counting
  on {Si}3n4 photonic circuits,'' {\em Appl. Phys. Lett.}~{\bf 102},  051101
  (Feb. 2013).

\bibitem{schuck_waveguide_2013}
Schuck, C., Pernice, W. H.~P., and Tang, H.~X., ``Waveguide integrated low
  noise {NbTiN} nanowire single-photon detectors with milli-{Hz} dark count
  rate,'' {\em Scientific Reports}~{\bf 3},  1893 (May 2013).

\bibitem{schuck_quantum_2016}
Schuck, C., Guo, X., Fan, L., Ma, X., Poot, M., and Tang, H.~X., ``Quantum
  interference in heterogeneous superconducting-photonic circuits on a silicon
  chip,'' {\em Nature Communications}~{\bf 7},  ncomms10352 (Jan. 2016).

\bibitem{beyer_waveguide-coupled_2015}
Beyer, A., Briggs, R., Marsili, F., Cohen, J.~D., Meenehan, S.~M., Painter,
  O.~J., and Shaw, M., ``Waveguide-{Coupled} {Superconducting} {Nanowire}
  {Single}-{Photon} {Detectors},'' in [{\em {CLEO}: 2015 (2015), paper
  {STh}1I.2}{\nolinebreak\hspace{0.1em}]},   STh1I.2, Optical Society of
  America (May 2015).

\bibitem{shainline_room-temperature-deposited_2017}
Shainline, J.~M., Buckley, S.~M., Nader, N., Gentry, C.~M., Cossel, K.~C.,
  Cleary, J.~W., Popovi{\'c}, M., Newbury, N.~R., Nam, S.~W., and Mirin, R.~P.,
  ``Room-temperature-deposited dielectrics and superconductors for integrated
  photonics,'' {\em Opt. Express}~{\bf 25},  10322--10334 (May 2017).

\bibitem{rath_superconducting_2015}
Rath, P., Kahl, O., Ferrari, S., Sproll, F., Lewes-Malandrakis, G., Brink, D.,
  Ilin, K., Siegel, M., Nebel, C., and Pernice, W., ``Superconducting
  single-photon detectors integrated with diamond nanophotonic circuits,'' {\em
  Light Sci Appl}~{\bf 4},  e338 (Oct. 2015).

\bibitem{atikian_superconducting_2014}
Atikian, H.~A., Eftekharian, A., Jafari~Salim, A., Burek, M.~J., Choy, J.~T.,
  Hamed~Majedi, A., and Lon{\v c}ar, M., ``Superconducting nanowire single
  photon detector on diamond,'' {\em Appl. Phys. Lett.}~{\bf 104},  122602
  (Mar. 2014).

\bibitem{heeres_-chip_2010}
Heeres, R.~W., Dorenbos, S.~N., Koene, B., Solomon, G.~S., Kouwenhoven, L.~P.,
  and Zwiller, V., ``On-{Chip} {Single} {Plasmon} {Detection},'' {\em Nano
  Lett.}~{\bf 10},  661--664 (Feb. 2010).

\bibitem{heeres_quantum_2013}
Heeres, R.~W., Kouwenhoven, L.~P., and Zwiller, V., ``Quantum interference in
  plasmonic circuits,'' {\em Nat Nano}~{\bf 8},  719--722 (Oct. 2013).

\bibitem{gerrits_-chip_2011}
Gerrits, T., Thomas-Peter, N., Gates, J.~C., Lita, A.~E., Metcalf, B.~J.,
  Calkins, B., Tomlin, N.~A., Fox, A.~E., Lamas~Linares, A., Spring, J.~B.,
  Langford, N.~K., Mirin, R.~P., Smith, P. G.~R., Walmsley, I.~A., and Nam,
  S.~W., ``On-chip, photon-number-resolving, telecommunication-band detectors
  for scalable photonic information processing,'' {\em Phys. Rev. A}~{\bf 84},
  060301 (Dec. 2011).

\bibitem{calkins_high_2013}
Calkins, B., Mennea, P.~L., Lita, A.~E., Metcalf, B.~J., Kolthammer, W.~S.,
  Lamas-Linares, A., Spring, J.~B., Humphreys, P.~C., Mirin, R.~P., Gates,
  J.~C., Smith, P. G.~R., Walmsley, I.~A., Gerrits, T., and Nam, S.~W., ``High
  quantum-efficiency photon-number-resolving detector for photonic on-chip
  information processing,'' {\em Opt. Express}~{\bf 21},  22657--22670 (Sept.
  2013).

\bibitem{marsili_detecting_2013}
Marsili, F., Verma, V.~B., Stern, J.~A., Harrington, S., Lita, A.~E., Gerrits,
  T., Vayshenker, I., Baek, B., Shaw, M.~D., Mirin, R.~P., and Nam, S.~W.,
  ``Detecting single infrared photons with 93\% system efficiency,'' {\em Nat
  Photon}~{\bf 7},  210--214 (Mar. 2013).

\bibitem{lita_counting_2008}
Lita, A.~E., Miller, A.~J., and Nam, S.~W., ``Counting near-infrared
  single-photons with 95\% efficiency,'' {\em Opt. Express}~{\bf 16},
  3032--3040 (Mar. 2008).

\bibitem{edwards_temperature-dependent_1984}
Edwards, G.~J. and Lawrence, M., ``A temperature-dependent dispersion equation
  for congruently grown lithium niobate,'' {\em Opt Quant Electron}~{\bf 16},
  373--375 (July 1984).

\bibitem{jundt_temperature-dependent_1997}
Jundt, D.~H., ``Temperature-dependent {Sellmeier} equation for the index of
  refraction, n$_{\textrm{e}}$, in congruent lithium niobate,'' {\em Opt.
  Lett., OL}~{\bf 22},  1553--1555 (Oct. 1997).

\bibitem{regener_loss_1985}
Regener, R. and Sohler, W., ``Loss in low-finesse {Ti}:{LiNbO}3 optical
  waveguide resonators,'' {\em Appl. Phys. B}~{\bf 36},  143--147 (Mar. 1985).

\end{thebibliography}
\bibliographystyle{spiebib} 

\end{document}